\documentclass[prl,twocolumn,showpacs]{revtex4}
\usepackage{amsmath,amssymb}
\newcount\minute
\newcount\hour
\newcount\hourMins
\def\now{
  \minute=\time
  \hour=\time \divide \hour by 60
  \hourMins=\hour \multiply\hourMins by 60
  \advance\minute by -\hourMins
  \zeroPadTwo{\the\hour}:\zeroPadTwo{\the\minute}
}
\def\zeroPadTwo#1%
{%
\ifnum #1<10 0\fi
#1%
}

\begin{document}

\title{\textsc{Transport  in a Dissipative Luttinger Liquid}}

\author{Zoran Ristivojevic and Thomas Nattermann
}

\affiliation{Institut f\"ur Theoretische Physik, Universit\"at zu
K\"oln, Z\"ulpicher Str. 77, 50937 K\"oln, Germany}

\date{\today,\now}

\begin{abstract}
We study theoretically the transport
through a single impurity in a one-channel
Luttinger liquid coupled to a dissipative
(ohmic) bath . For non-zero dissipation
$\eta$ the weak link is always a relevant
perturbation which suppresses transport
strongly. At zero temperature the current
voltage relation of the link is $I\sim
\exp(-E_0/eV)$ where $E_0\sim\eta/\kappa$
and $\kappa$ denotes the compressibility.
At non-zero temperature $T$ the linear
conductance is proportional to
$\exp(-\sqrt{{\cal C}E_0/k_BT})$. The
decay of Friedel oscillation saturates for
distance larger than $L_{\eta}\sim 1/\eta
$ from the impurity.

\end{abstract}

\maketitle

\emph{Introduction}. One dimensional
interacting  electron systems  behave
quite differently from their higher
dimensional counterparts. Whereas in two
and three dimensions interacting electrons
form a Fermi fluid characterized by the
same quantum numbers as free electrons
\cite{Noziere}, the excitations in one
dimensional electron systems  are of
collective nature - essentially because
the electrons cannot avoid each other. To
create a single electron requires the
excitation of an infinite number of these
collective phonon-like excitations
(plasmons). In this case electrons form a
new state of matter: the Luttinger liquid
(LL) \cite{FiGl96,Gi03}.

Renewed interest in Luttinger liquids
arises from progress in manufacturing
narrow quantum wires with a few or a
single conducting channel. Examples are
carbon nanotubes \cite{CuZe04},
polydiacetylen \cite{AlLe04}, quantum Hall
edges \cite{KaSt00} and semiconductor
cleave edge quantum wires \cite{AuYa02}.
More recently LL have been realized in the
core of a screw dislocation in \emph{hcp}
$^4$He crystals \cite{BoKu}.

A clean LL is  a marginal supersolid
exhibiting both a power law decay of
superfluid and $2k_F$ {density}
correlations. It is described by a single
parameter $K$ which is a measure of the
short range interaction between the
electrons, with $K>1$ for attractive and
$K<1$ for repulsive interaction,
respectively. In a contact free
measurement a clean Luttinger liquid has a
finite ac conductance $G(\omega)=Ke^2/h$,
provided the  system size $L$ is smaller
than $L_{\omega}=v/\omega$
 \cite{ApRi82,OgFu94,remark},  $v$ denotes the
 plasmon
 velocity. In the opposite limit
 $L\gg v/\omega$ the system has a complex
 conductivity $\sigma(\omega)\approx
 2vG/(-i\omega+\varepsilon)$, similar to a superconductor.

An isolated impurity leaves  the linear
zero temperature conductance of the LL
unchanged, provided the interaction is
attractive. This result is independent of
the strength of the impurity.  For
repulsive interaction, on the other hand,
the conductance vanishes at zero
temperature as $G\sim V^{2/K-2}$, where
$V$ is the applied voltage
\cite{Kane+92b,MaGla93}. At non-zero
temperatures $T$ the conductance $G\sim
T^{2/K-2}$ remains finite even for $V\to
0$.

In realistic  systems electrons in a LL
couple to other degrees of freedom, e.g.
to phonons or to electrons in nearby
gates. These degrees of freedom act as a
dissipative bath \cite{Caldera_Leggett}.
In particular it  has been shown recently
that the coupling of a clean LL to gate
electrons results in an ohmic dissipation
of the electrons in the LL
\cite{Cazalilla+06}. Dissipation
introduces a further length scale
$L_{\eta}\approx 1/(K\eta)$ where $\eta$
denotes the dissipation strength. On
scales $L>L_{\eta}$  plasmons become
diffusive and  displacement fluctuations
 are strongly suppressed restoring translational
 long range order (Wigner crystal).  If
$L_{\omega}\gg L_{\eta}$, i.e. $ Kv\eta\gg
\omega$, the conductivity
$\sigma=2e^2/(h\eta)$ is finite which is
paralleled by diverging superfluid
fluctuations.

The question arises what is the influence
of dissipation on the transport properties
of disordered LL? This question will be
addressed in the present paper. In
particular we study  the influence of weak
dissipation on the tunneling of electrons
through a  single impurity. It turns out
that the dissipation strongly suppresses
tunneling and hence changes the power law
dependence of the conductance on $V$ and
$T$ to an exponential one.

\bigskip

\emph{The model}. We consider spinless
electrons in a one dimensional wire which
are coupled to a dissipative bath. In the
center of the wire is a single impurity of
arbitrary strength. Following Haldane
\cite{Haldane} we rewrite the charge
density $\rho(x)$ as
\begin{equation}\label{eq:density}
    \rho = \pi^{-1}(k_F + \partial_x\varphi)
           [1 + \cos(2\varphi + 2 k_F x) + \ldots].
\end{equation}
where  $\varphi(x)$ denotes a bosonic
(plasmon) displacement field. The
Euclidean action of the system then reads
\cite{Gi03}
\begin{align}\label{eq:full_action1}
&\frac{S}{\hbar} = \frac{1}{2\pi K }
    \int\limits_{-L/2}^{L/2} \int
    \limits_{-\hbar /2k_BT}^{\hbar /2k_BT}dx
    d\tau
    \Bigg\{\frac{1}{v}(\partial_{{\tau}}\varphi)^{2}+
    v(\partial_{x}\varphi
    )^{2}
    \\
  &-W(\varphi)\delta(x)+\frac{\pi K \eta}{2}
   \int\limits_{-\hbar/2k_BT}^{\hbar /2k_BT} \frac{d\tau'
(\varphi(x,\tau)-\varphi(x,\tau')^2}
{[\hbar\beta\sin(\pi(\tau-\tau')/\hbar
\beta)]^2}\Bigg\}. \nonumber
\end{align}
 For $K\ll 1$
(\ref{eq:full_action1}) describes also a
charge or spin density wave, $K\to 0$
corresponds to the classical limit.
$W(\varphi)$ is the impurity potential
which is a periodic function of
periodicity $\pi$. The last part of
(\ref{eq:full_action1})
 describes ohmic dissipation \cite{Caldera_Leggett}.
It was  derived  from a pure Luttinger
liquid coupled electrostatically to a
metallic gate in \cite{Cazalilla+06} where
it was found
 that such a coupling
is only relevant for $K<K_{\eta}=1/2$.
This fact can be taken into account by
writing $\eta=\eta_0\xi_{\eta}^{-1}$ where
$\xi_{\eta}$ denotes the
Kosterlitz-Thouless correlation length
diverging at the transition $K\to
K_{\eta}-0$ and $\eta_0$ is a
dimensionless coupling constant. Since
there may be other sources of dissipation
which  survive also for $K>1/2$ in the
following we will assume that $\eta>0$.

\emph{Effective action}. In treating the
influence of  the impurity  we first
consider the case of zero temperature.
After Fourier transforming
$\varphi(x,\tau)=\int_{\omega}\int_k
\varphi_{k,\omega}e^{ikx+i\omega\tau}$
where
$\int_{\omega}\equiv\int\frac{d\omega}{2\pi}$
etc. the harmonic action reads
\begin{eqnarray}
  \frac{S_0}{\hbar}= \frac{1}{2\pi K }
    \int_{\omega} \int_k\left\{\frac{1}{v}\omega^2+vk^2+
K\eta|\omega|\right\}
    |\varphi_{k,\omega}|^{2}.
\label{eq:full_action}
\end{eqnarray}

The last term in (\ref{eq:full_action})
describes the (weak) damping of the
plasmons of complex frequency
$\omega_P\approx v(k+iK\eta/2)$. The upper
cut-off for momentum and frequency are
$\Lambda$ and $\omega_c$ , respectively.
$\omega_c^{-1}=\textrm{min}\{v\Lambda,
E_{\textrm{diss}}/\hbar\}$ where
$E_{\textrm{diss}}$ corresponds to the
maximal energy which can be dissipated by
the bath.  For simplicity we will assume
that they both are of the same order of
magnitude. We integrate now in the
standard procedure  over the degrees of
freedom outside of the impurity. The
resulting effective impurity action now
reads
\begin{eqnarray}
  \frac{S}{\hbar}\approx\frac{1}{\pi K }
    \int_{\omega} \left\{\omega^2+
Kv\eta|\omega|\right\}^{1/2}
    |\phi_{\omega}|^{2}-\int d\tau W(\phi) \label{eq:effective_action}
\end{eqnarray}
where $\phi_{\omega}=\int d\tau
\phi(\tau)e^{-i\tau\omega}$ and
$\phi(\tau)\equiv\varphi(x=0,\tau)$. As
can be seen from
(\ref{eq:effective_action}) for low
frequencies, $\omega<\omega_{\eta}=
v/L_{\eta}$, the behavior of the system is
controlled by the dissipation. For very
large values of $\omega$ and sufficiently
small dissipation, $\eta\ll \Lambda/K$,
the effective action includes also a
contribution $\sim
\int_{\omega}\omega^2|\phi_{\omega}|^{2}$
\cite{Gi03}.

\bigskip
\emph{Weak impurity}. Since $W(\phi)$ is a
periodic function we can decompose it in a
Fourier series
   $ W(\phi) = \sum_{n=1}^{\infty} w_n\cos
    (2n\phi).$
For weak impurity strength we can
calculate the renormalization of $W(\phi)$
perturbatively. This gives for the flow
equations of $w_n$
\begin{equation}\label{eq:impurity_strength}
    \frac{dw_n}{dl}=\left(1-\frac{n^2K}{\sqrt{1+\eta
    Ke^{l}/\Lambda}}\right)w_n.
\end{equation}
Since the effective action depends in a
non-analytic way on the frequency there is
no renormalization of $K$ and $\eta$.

For $\eta=0$ we recover from
(\ref{eq:impurity_strength}) the result of
Kane and Fisher \cite{Kane+92b} namely
that the impurity is a relevant
perturbation for repulsive interaction and
irrelevant for attractive interaction. As
soon as the dissipation is switched on the
downward renormalization saturates at a
length scale $\Lambda^{-1}e^{l^*}\approx
L_{\eta}$. Thus the impurity is a relevant
perturbation even for attractive
interaction, provided the dissipation
remains finite in this region
\cite{renormailzation_of_eta}.

\emph{Strong impurity.} In the case of a
strong impurity potential the main
contribution to the partition function
will come from configurations
$\phi(\tau)=n\pi$ with $n$ integer,
interrupted by kinks which connect these
pieces. These multi-kinks configurations
are saddle points $\phi_s(\tau)$ of the
effective action
(\ref{eq:effective_action}).  For very
large impurity strength kinks are narrow
and hence are characterized by the high
frequency limit of
(\ref{eq:effective_action}). A solution
with $n$ kinks and $n$ antikinks can then
be written as
\begin{equation}\label{}
    \phi_s(\tau)\approx\pi\sum_{i=1}^{2n}
    \epsilon_i\Theta(\tau-\tau_i),
    \,\,\,\epsilon_i=\pm1,\,\,\,\,\sum_{i=1}^{2n}
    \epsilon_i=0
\end{equation}
where $\Theta(\tau)$ is the  Heaviside
step function and $\tau_i$ the kink
positions. In frequency space the saddle
point configuration can be written as
\begin{equation}\label{}
    \phi_{s,\omega}\approx\frac{i\pi}{\omega}
    \sum_{i=1}^{2n}\epsilon_i
    e^{i\omega\tau_i}.
\end{equation}
The partition function then
reads
\begin{eqnarray} \label{partition_function}
 Z &=&
\sum_{n=0}^{\infty}\sum_{\epsilon_j=\pm1}\frac{t^{2n}}{(2n)!}\int
d\tau_1...d\tau_{2n} e^{\frac{2}{K}
\sum\limits_{i<j}\epsilon_i\epsilon_jf(\tau_i-\tau_j)}
  \end{eqnarray}
Here $t=e^{-S_{\textrm{kink}}/\hbar}$ is
the tunneling transparency,
$S_{\textrm{kink}}\sim \sqrt{w_1}$ denotes
the action of an isolated kink and
$f(\tau)$ is the kink interaction
\begin{equation}\label{eq:kink-interaction1}
    f(\tau)=\int_0^{\omega_c}{d\omega}\frac{\sqrt{\omega^2+
    v
    K\eta|\omega|}}{\omega^2}\left(1-\cos(\omega
    \tau)\right)
\end{equation}
Using a soft cut-off $\sim
e^{-\omega/\omega_c}$ in
(\ref{eq:kink-interaction1}) one obtains
in the limit of  $\omega_c\tau\gg 1$ as an
interpolating expression (up to a
constant)
\begin{equation}\label{}
    f(\tau)\approx \ln\omega_c\tau
    +(2\pi v\eta K\tau)^{1/2}.
\end{equation}
The logarithmic interaction prevails for
small kink-anti-kink distances $\tau$
whereas for larger $\tau$ the interaction
exhibits power law behavior.

A strong impurity can alternatively be
considered as a weak link between the two
half-wires $x<0$ and $x>0$, respectively.
It is therefore convenient to rewrite the
partition function
(\ref{partition_function}) in a form in
which $t$ denotes the strength of the
non-linear terms. This can be done in the
standard way \cite{Schmid83,FuNa93} with
the result $  Z=\int {\cal D}\theta
    e^{-S_{\textrm{eff}}\{\theta\}/\hbar}$
\begin{eqnarray}\label{}
\frac{S_{\textrm{eff}}}{\hbar}=\frac{1}{\pi}\int_{\omega}
\frac{K\omega^2}{\sqrt{\omega^2+v\eta K
|\omega|}}|\theta_{\omega}|^2-2t\int d\tau
\cos 2\theta(\tau).\nonumber
\end{eqnarray}

 We
can now consider the renormalization of
$t$ which is given by
\begin{equation}\label{eq:impurity_strength2}
    \frac{dt}{dl}=\left(1-K^{-1}{\sqrt{1+\eta
    Ke^{l}/\Lambda}}\right)t.
\end{equation}
 Again
$\eta\equiv 0$  reproduces the known
result \cite{Kane+92b} that the weak link
is renormalized to zero for $K<1$. In the
dissipative case the  link is renormalized
to zero for all $K$ provided $\eta$
remains finite. Thus from
(\ref{eq:impurity_strength}) and
(\ref{eq:impurity_strength2}) we conclude
that the impurity is always a relevant
perturbation
\cite{renormailzation_of_eta}.

 \emph{Transport}.
 Next we calculate  the current $I$ through the
 impurity. The tunneling rate $\Gamma$ and hence the
current follows from \cite{LaOv84}
\begin{equation}\label{eq:Gamma}
    \hbar\Gamma=2(k_BT)^{-1}{{ \textrm{Im}}}\ln
    Z(V),\,\,\,\,\,\,\,Z=Z_0+iZ_1
\end{equation}
where $Z(V)$ denotes the partition
function in the presence of an external
voltage $V$. $Z_0$ includes the (stable)
fluctuations of the $\phi$-field close to
a local minimum $\phi=n\pi$ whereas $Z_1$
includes the voltage driven (unstable)
fluctuations which connect neighboring
minima $\phi=n\pi\to (n+1)\pi$. Since the
the strength of the impurity is
renormalized to large values we can use
the saddle point approximation employed in
the previous paragraph for the calculation
of $Z$. To obtain the saddle point action
it is  sufficient to consider a
configuration consisting of  a pair of
kinks with distance $\tau$
\begin{equation}\label{eq:double-kink-action}
    \frac{S(\tau)}{\hbar}=\frac{2}{K}f(\tau)+2S_{\textrm{kink}}-
     \frac{eV}{\hbar}\tau-\ln \tau.
\end{equation}
Here we have taken into account that the
voltage creates a term $-e\int
d\tau{\phi}V/{\pi}$ in $S$
\cite{voltage}.
  The last term is of entropic
origin and describes fluctuations in the
kink distance. The saddle point follows
then from
\begin{equation}\label{}
  \frac{2}{K}f'(\tau)-\frac{e V}{\hbar}-\frac{1}{\tau}=0.
\end{equation}
This gives in for larger voltages for the
saddle point $\tau_c\approx
(2/K-1)\hbar/(eV)$ and hence for the
current
\begin{equation}\label{eq:Kane}
    I\sim e^{-S(\tau_c)}\sim t^2\left({e
    V}/{\hbar\omega_c}\right)^{2/K-1},\,\,\,\,
    eV\ll \hbar\omega_c
\end{equation}
i.e. we reproduce the result of Kane and
Fisher \cite{Kane+92b} for the conductance
$G\sim \left({e
    V}/{\hbar\omega_c}\right)^{2/K-2}$.
On the contrary, in the dissipative limit
$v\eta K\tau\gg 1$ we get for the saddle
point $\tau_c\approx {2\pi\eta
v\hbar^2}/({e^2 V^2}K)$ and hence for $e
V\kappa\ll \eta$

\begin{equation}\label{eq:I-V}
    I\approx Ae^{-S(\tau_c)/\hbar}=
    A(V) \,t^2
    \exp\Big\{-\frac{4\eta}
    {\kappa eV}\Big\}  .
\end{equation}
Here we introduced the compressibility
$\kappa =K/(\pi v \hbar)$. In the case of
non-interacting electrons $\kappa$
coincides with the density of states
\cite{qualitative}. As expected, the
dissipation reduces the current with
respect to the Kane-Fisher result
strongly. This means that the tunneling
density of states is suppressed more than
a power law \cite{density_of_states}. The
cross-over between (\ref{eq:I-V}) and
(\ref{eq:Kane}) happens at
$eV/\hbar\approx \eta v$. An independent
calculation using Fermi's ''golden rule''
 gives for the pre-factor
$A(V)\sim (\eta/\kappa (eV)^3)^{1/2}$
\cite{Zoran}.

 Next we consider  finite
 temperatures. In this case the extension
 of the $\tau$-axis is finite. Hence there
 will be a cross-over from (\ref{eq:I-V}) to a new
 behavior when the instanton
 hits the boundary, i.e. if $\tau_c\approx
 \hbar/ k_BT$. In the dissipation free limit
 this results in a conductance $G\sim
 (k_BT/\hbar\omega_c)^{2/K-2}$. In
 the dissipative case this $ eV$ to $k_BT$ dependence cross-over
 happens for $eV\approx\sqrt{2\eta k_BT/\kappa}$.
 For larger $T$ the current is given by
\begin{equation}\label{eq:I-T}
    I\approx B(T) \exp\left\{-\sqrt{\frac{{\cal C}\eta }{\kappa k_BT}}\right\}V,\,\,\,\,\,
    \kappa k_BT\ll {\cal C}\eta.
\end{equation}
The calculation does not allow a precise
determination of the numerical constant
${\cal C}$ (${\cal C}=8$ from the present
approach). The result resembles variable
range hopping but results here from
sequential tunneling through individual
impurities. An independent calculation
gives $B(T)\sim T^2\left(\kappa
T/\eta\right)^{1/4}$ \cite{Zoran}.

A simple way to reproduce (\ref{eq:I-T})
follows from integrating the flow equation
for the transparency
(\ref{eq:impurity_strength2}) and
truncating the RG flow at the thermal de
Broglie wave length $L_T=\hbar v /k_BT$ of
the plasmons \cite{Gi03}. This gives an
effective tunneling transparency
\begin{equation}\label{eq:t_eff}
    t_{\textrm{eff}}\approx t_0\frac{\hbar
    v\Lambda}{k_BT}
    \exp\left\{-\sqrt{\frac{4\eta}{\pi k_BT\kappa}}\right\}.
\end{equation}
The total current is then proportional to
$t_{\textrm{eff}}^2$ which agrees with
(\ref{eq:I-T}) apart from the value of
${\cal C}$.

 \emph{Friedel
oscillations.} Finally we calculate the
charge density oscillations close to the
defect.  In the absence of dissipation
they decay $\langle\rho(x)\rangle \sim
\cos (2k_Fx)|x|^{-\alpha} $ where
$\alpha=K$ for $K<1$ and $\alpha =2K-1 $
for $K>1$ \cite{Kane+92a}. To determine
the averaged charge density we write
\begin{equation}\label{}
\pi\langle\rho(x)\rangle-k_F=\langle
\partial_x\varphi\rangle
+k_F\cos(2k_Fx)e^{-2\langle\varphi^2(x)\rangle}.
\end{equation}
The disorder average in the first term
vanishes. To calculate the second term we
remark that in the presence of dissipation
the impurity is always relevant, its
strength (\ref{eq:impurity_strength})
growths under renormalization. To
determine the large scale behavior of
$\langle\rho(x)\rangle$ it is therefore
justified to assume that $\varphi$ is
fixed at the defect. The fluctuation
$\langle\varphi^2(x)\rangle$ is therefore
identical to
$\frac{1}{2}\langle(\varphi(x)-\varphi(-x))^2\rangle$.
From this one finds  for $x\ll L_{\eta}$
$\langle\rho(x)\rangle\sim\cos(2k_Fx)
|x|^{-K}$, i.e. the power law decay well
known from Luttinger liquids. However for
$|x|\gg L_{\eta}$ the amplitude behaves
like
\begin{equation}\label{}
    \langle\rho(|x|\gg L_{\eta}
    )\rangle\approx\frac{\pi^{-1}k_F\cos(2k_Fx)}{( 2\sqrt{y^2+y}
    +2y+1)^K}\left(1+\frac{KL_{\eta}}{|x|}\right)
    ,
\end{equation}
with $
    y={\omega_c}/(v{\eta K})\equiv \Lambda L_{\eta}$.

    \emph{Many impurities}. So far we
    considered the influence of dissipation on the transport through
    a single impurity. In the case of many
    randomly distributed impurities
     with mean spacing $a$ we
    expect no influence of dissipation as
    long as $a\ll L_{\eta}$. This agrees
    with our previous findings on Gaussian
    disorder \cite{MaNaRo04} where  weak
    dissipation has to be taken into account
    to guarantee energy conservation. In
    this case
    one finds variable range hopping with
    only a weak dissipation dependence if
    one restricts oneself  to
    exponential accuracy. For $L_{\eta}\ll
    a$ it is suggesting that if $L_T\ll a$
    the different impurities are decoupled
    and one observes single impurity
    behavior.

\emph{Conclusions.} We have shown that the
inclusion of weak ohmic dissipation has a
dramatic effect on the transport
properties of a LL through a single
impurity. The voltage and temperature
dependence of the conductance is reduced
from power laws in the dissipation free
case to an exponential dependence, eqs.
(\ref{eq:I-V}), (\ref{eq:I-T}).

We thank Bernd Rosenow for helpful
discussions. This work was supported by
the Deutsche Forschungsgemeinschaft
through NA 222/5-2 (Z.R.) and  SFB 608/D4
(T.N.).


\begin{thebibliography}{99}
\bibitem{remark} Note that the conditions
$L\ll v/\omega$, $L\gg v/\omega$  do not
express the relevance of the leads but the
conditions under which one finds a finite
conductance and conductivity,
respectively.
\bibitem{renormailzation_of_eta} If the
dissipation results from the coupling to
gate electrons, as considered in
\cite{Cazalilla+06}, $\eta$ renormalizes
to zero for $K>1/2$ and hence the impurity
potential renormalizes to zero for $K>1$.
\bibitem{voltage} For not too long wires,
$L\lesssim \hbar v/(KeV)$ the voltage drop
due to the dissipation in the two half
wires can be ignored and $V$ is the total
voltage applied at the wire.
\bibitem{qualitative} In the noninteracting case, $K=1$, eq.
(\ref{eq:I-V}) can be obtained from the
following argument. The probability that
an electron tunnels over $r$ distance is
proportional to $W(r)\sim
e^{-2r/L_{\eta}}$. To find the minimal $r$
the number of states which can be reached
has to be at least of the order one, i.e.
$\kappa eV r\approx 1$. Inserting $r$ into
$W(r)$ gives (\ref{eq:I-V}) apart from a
numerical factor.
\bibitem{density_of_states} Rewriting
$I\sim t^2 \int_0^{eV}\rho(E)\rho(eV-E)$
where $\rho(E)$ denotes the tunneling
density of states we conclude from
(\ref{eq:I-V}) $\rho(E)\sim e^{-{\cal
C}'\eta /(\kappa E)}$, i.e. the tunneling
density of states is strongly suppressed
by the dissipation in comparison to the
power law found in \cite{Kane+92b}.
\end{thebibliography}
\end{document}